# Personalized Email Community Detection using Collaborative Similarity Measure

[1]Waqas Nawaz*, [2]Yongkoo Han, [3]Kifayat-Ullah Khan, [4]Young-Koo Lee
Department of Computer Engineering, Kyung Hee University
Yongin-si, Gyeonggi-do 446-701, Republic of Korea
{wicky786[1], ykhan[2], yklee[4]}@khu.ac.kr, kualizai@hotmail.com[3]

*Abstract*
Email service providers have employed many email classification and prioritization systems over the last decade to improve their services. In order to assist email services, we propose a personalized email community detection method to discover the groupings of email users based on their structural and semantic intimacy. We extract the personalized social graph from a set of emails by uniquely leveraging each node with communication behavior. Subsequently, collaborative similarity measure (CSM) based intra-graph clustering approach detects personalized communities. The empirical analysis shows effectiveness of the resultant communities in terms of evaluation measures, i.e. density, entropy and f-measure. Moreover, email strainer, dynamic group prediction, and fraudulent account detection are suggested as the potential applications from both the service provider and user's point of view.

**Keywords**: *Graph Mining, Personalized Clustering, Data Mining, Unsupervised Classification, Personalization, Similarity Measure, Collaborative, Density, Entropy, F-Measure, Email Network*

## 1. Introduction

Social community detection is an interesting research area [1]. It can be used for classification of emails, discovery of prominent users, and highlighting abnormal activities inside the network [2] [3] [4] [5]. However, majority approaches analyze the entire email network. Practically, it is unrealistic to acquire multi-user email data due to privacy issues. It is also inefficient to analyze an entire email network for personalized applications. These approaches are inappropriate for investigating individual behavior and relationships based on personal emails within the social network.

This paper presents an effective social community detection method over a personalized email network, which is solely based on emails owned by a particular user. Personal emails reflect the directed social interaction among individuals. This kind of interaction is regarded as an undirected weighted graph (UW-Graph) where the directional information is aggregated on each edge. Each user, i.e. either sender or receiver, is represented by a node and an edge reflects shared emails where frequency is associated as an edge weight. The personalization can leads us towards the best approximation of an individual's behavioral activities. The first phase of our proposed method extracts the communication patterns of interest (CPI) as informative features to describe the communication behavior of each user. Subsequently, the second phase detects user communities via an intra-graph clustering method which contemplates structural and semantic aspects together. The semantic resemblance among individuals is achieved through CSM. We validate the effectiveness of the proposed technique on real email dataset in terms of various evaluation measures (i.e. density, entropy, and f-score).

## 2. Related Work

In the recent past, many researchers have analyzed entire email networks to study influential users, email prioritization, community structures, and organizational hierarchies based on statistical or structural analysis. We explore a few recent approaches sequentially with respect to their relevance to our work.

Among the early efforts in email classification based on statistical analysis, Steve Martin et al. [6] have highlighted the importance of outgoing email information among individuals rather than incoming emails. They have also analyzed the contribution of the features extracted from emails to email classification. Moreover, the characterization of emails and community identification is also achieved by L. Johansen et al. [7] based on the flow and frequency of email interaction among users. However, the threshold value may vary for each user. It is very hard to generalize an email frequency for effective user grouping. Few techniques are intended to predict the organizational structure solely based on interaction and relationship among individuals through emails [8] [9] [10]. These techniques are inspired by sent & received emails or page rank. Unfortunately, this kind of analysis tends to determine the authority level of individuals based on their importance in the communication network rather than grouping centered on the nature of user interaction.

On the other hand, Shinjae Yoo et al. have utilized the topological structure of the email network for non-Spam Email prioritization. Hanhe Lin [11] has identified hidden relationships in the email network to highlight mutual private communication. Usually, private relationships cannot lead us to a grouping of users with similar email usage patterns or behaviors. In contrast to hidden relationships, social network analysis has the ability to understand the community structure. G. Wilson [12] has used SNA in conjunction with an evolutionary method, i.e. Genetic Algorithm, to identify key email users. In contrast to various approaches, our intention in this paper is to discover communities or group of individuals based on similar email communication patterns and email network structures. Our approach is to solely consume personal emails instead of entire email networks. In subsequent sections, a way of capturing both aspects in the presence of personalized information and a partitioning strategy has been discussed in detail.

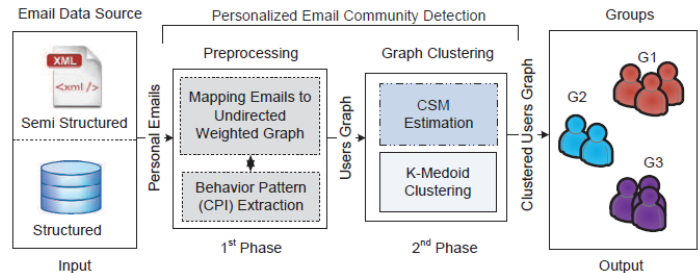

**Figure 1.** System Architecture for Social Community Detection from Personal Emails

## 3. Personalized Email Community Detection

The ultimate goal of Personalized Email Community Detection is to portray the grouping of individuals with similar neighborhood and communication behavior. The architecture of the proposed method with two mutually exclusive, sequentially dependent phases is given in Figure 1. Initially, email information is acquired from a raw data source which consists of electronic communication among people. In the preprocessing phase, the interactions among the users are represented in the form of an UW-Graph. A set of attributes are attached with each node to reflect the user's communication pattern or behavior. In the second phase, the groups of individuals are identified based on their overlapping communication interests using an intra-graph clustering approach [13] which is simple and effective in this context. It has the capability to capture both topological and semantic aspects concurrently during the clustering process.

### 3.1. Preprocessing Phase

At the beginning, email information owned by a particular user is provided as an input to the system. This information is directly extracted from personal emails due to privacy, unavailability of multi-user email data sources, and intensive computations. Moreover, it is represented as a UW-Graph describing behavioral patterns for further analysis, unlike the case with prior techniques. In the graph domain, each user is represented by one vertex and the intensity of emails exchanged is reflected by link weights. We do not care about the directional aspect of the real email network because we are not dealing with the flow of information in the network. Our focus is to capture an individual's behavior with regard to information exchange. The graphical description of the transition or mapping strategy from personal directed email network to a UW-Graph is given in Figure 2.

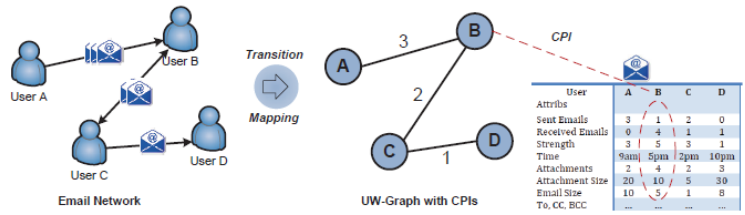

**Figure 2.** Personal Emails towards UW-Graph

### 3.2. Community Detection using CSM

In order to discover underlying user community structures from this personalized social graph, a graph partitioning approach, i.e. intra graph clustering [13], is exploited. It has the capability to group all the multi-attributed nodes. This partitioning strategy considers both the topological structure and semantic resemblance (or CPI) among nodes. Ultimately, we obtain groups of email users based on their communication interactions and behavior.

### 4. Experiments

The proposed method has been implemented in Java[1]. We have considered the Enron email dataset[2]. A single user's, Sally Beck, email network is exploited from this data-set for all the experiments in this paper. All the users with strong relationships due to email exchange are grouped together by employing a density function. Similarly, the quality of the resultant communities is determined by semantic relevance among users based upon their attributed nature. Accordingly, the structural and contextual quality of the communities has been analyzed in terms of Density and Entropy, respectively, by varying the number of communities as shown in Figure 3(a). The visual representation of four communities and statistical information is given in Figure 3(b) and Table 1 respectively.

**Table 1.** The statistical information [14] of the community structure in Figure 3(b)

| Parameters | Com-1 | Com-2 | Com-3 | Com-4 |
|---|---|---|---|---|
| Clus Coefficient | 0.056 | 0.274 | 0.492 | 0.355 |
| Centralization | 1.235 | 2.181 | 0.834 | 1.026 |
| Avg. Neighbors | 2.487 | 3.3 | 2.905 | 2.5 |
| Nodes | 39 | 10 | 21 | 18 |
| Network Density | 0.065 | 0.367 | 0.145 | 0.147 |

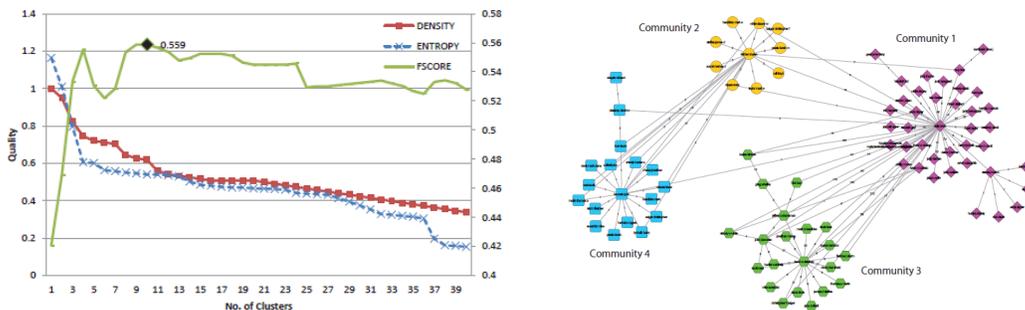

**Figure 3. (a)** No. of Communities vs. Quality: Sally-Beck's Personalized Communities **(b)** Sally-Beck's Four Personalized Communities.

Generally, community structures can be easily utilized to facilitate the security mechanism by identifying anomalous behavior and dynamically setting firewall rules. Automatic organization of incoming emails, based on some user-specified features, is also a generic approach, where content information, address books and communication patterns are considered to classify emails. Similarly, automatic email prioritization is another well-known type of email filter which can significantly reduce

---

[1] Proposed method's binaries will be provided on request (http://dke.khu.ac.kr/users/wicky786)
[2] http://www.edrm.net/resources/data-sets/edrm-enron-email-data-set-v2} (Last Access on Dec 15, 2012)

the amount of effort a user spends manually sorting through emails. Community information can also assist in automatic email mailing list generation which is specific to a particular user.

## 5. Conclusion

In this paper we have used a personalized email network under the assumption that only personal email data is available to discover user communities based on interactions and behavior patterns. Each resultant community retained a set of strongly connected users with similar behavioral patterns which are validated against density, entropy and f-measure. These results enable email service providers to offer their customers useful and interesting applications such as an email strainer, dynamic group prediction, and fraudulent account detection. Moreover, these personalized communities have the potential to approximate the nature and dynamics of an entire email network. This kind of approximation is subjective to a number of personal email accounts, which is beyond the scope of this paper.

## 6. Acknowledgments

This research was supported by the MSIP (Ministry of Science, ICT and Future Planning), Korea, under the ITRC (Information Technology Research Center) support program supervised by the NIPA (National IT Industry Promotion Agency) (NIPA-2013-(H0301-13-2001)).

## 7. References

[1] S. Papadopoulos, Y. Kompatsiaris, A. Vakali, and P. Spyridonos. Community detection in social media. Data Mining and Knowledge Discovery, 24(3):515-554, May 2012.

[2] J. Shetty and J. Adibi. Discovering important nodes through graph entropy the case of Enron email database. In Proceedings of the 3rd international workshop on Link discovery, LinkKDD '05, pages 74-81, New York, NY, USA, 2005. ACM.

[3] S. Yoo, Y. Yang, F. Lin, and I.-C. Moon. Mining social networks for personalized email prioritization. In Proceedings of the 15th ACM SIGKDD international conference on Knowledge discovery and data mining, KDD '09, pages 967-976, New York, NY, USA, 2009. ACM.

[4] J. L. J. H. Xin Jin, Cindy Xide Lin. Socialspamguard: A data mining-based spam detection system for social media networks. PVLDB, 4(12):1458-1461, 2011.

[5] Q. W. J. Z. H. G. J. W. J. F. Zhongying Zhao, Shengzhong Feng. Topic oriented community detection through social objects and link analysis in social networks. Knowl.-Based Syst., 26:164-173, 2012.

[6] S. Martin, A. Sewani, B. Nelson, K. Chen, and A. D. Joseph. Analyzing behavioral features for email classification. In Berkeley, CA: University of Caiifornia at Berkeley, 2005.

[7] L. Johansen, M. Rowell, K. Butler, and P. Mcdaniel. Email communities of interest, 2007.

[8] H. Yang, J. Luo, Y. Liu, M. Yin, and D. Cao. Discovering important nodes through comprehensive assessment theory on Enron email database. In Biomedical Engineering and Informatics (BMEI), 2010 3rd International Conference on, volume 7, pages 3041-3045, Oct. 2010.

[9] K. Yelupula and S. Ramaswamy. Social network analysis for email classification. In Proceedings of the 46th Annual Southeast Regional Conference on XX, ACM-SE 46, pages 469-474, New York, NY, USA, 2008. ACM.

[10] A. Timo eiev, V. Snasel, and J. Dvorsky. Social community detection in Enron corpus using h-index. In Applications of Digital Information and Web Technologies, 2008. ICADIWT 2008. First International Conference on the, pages 507-512, Aug. 2008.

[11] H. Lin. Predicting sensitive relationships from email corpus. In Genetic and Evolutionary Computing (ICGEC), 2010 Fourth International Conference on, pages 264-267, Dec. 2010.

[12] G. Wilson and W. Banzhaf. Discovery of email communication networks from the Enron corpus with a genetic algorithm using social network analysis, 2009.

[13] W. Nawaz, Y.-K. Lee, and S. Lee. Collaborative similarity measure for intra graph clustering. In SNSM, DASFAA, pages 204-215, 2012.

[14] Y. Assenov, F. Ram rez, S.-E. Schelhorn, T. Lengauer, and M. Albrecht. Computing topological parameters of biological networks. Bioinformatics, 24(2):282-284, 2008.